\documentclass[12pt,a4paper]{article}
\usepackage{graphicx}
\usepackage{times}
\newcommand{\mdot}{\ensuremath{\rm \dot M \;}}
\newcommand{\Mdot}{\ensuremath{\rm \dot M \;}}
\newcommand{\vinf}{\mbox{$\rm v_{\infty} \;$}}

\newcommand{\Lx}{$\log{ L_X/L_{bol}}$~}
\newcommand{\teff}{\mbox{$\rm T_{eff} \;$}}
\newcommand{\Teff}{\mbox{$\rm T_{eff} \;$}}
\newcommand{\logg}{\mbox{$\rm \log \, g \;$}}
\title{\bf Exploring the connection of weak winds and magnetic fields}
\author{Miriam Garcia$^{1,2}$, Francisco Najarro$^3$ and Artemio Herrero$^{1,2}$\\
\vspace{1cm}\\
\normalsize $^1$ Instituto de Astrof\'{\i}sica de Canarias, La Laguna, Spain \\ 
\normalsize $^2$ Universidad de La Laguna, La Laguna, Spain \\
\normalsize $^3$ Centro de Astrobiolog\'{\i}a (CSIC-INTA), Torrej\'on de Ardoz, Spain \\
}
\date{\mbox{}}
\begin{document}
\maketitle
\pagestyle{empty}
%
% WE REDEFINE THE plain LaTeX PAGESTYLE !!! 
% THIS PAGESTYLE WILL BE USED FOR THE FIRST PAGE ONLY !
%
\def\bull{\vrule height .9ex width .8ex depth -.1ex}
\makeatletter
\def\ps@plain{\let\@mkboth\gobbletwo
\def\@oddhead{}\def\@oddfoot{\hfil\tiny\bull\quad
``The multi-wavelength view of hot, massive stars''; 39$^{\rm th}$ Li\`ege Int.\ Astroph.\ Coll., 12-16 July 2010 \quad\bull}%
\def\@evenhead{}\let\@evenfoot\@oddfoot}
\makeatother
%
% AND DEFINE OUR MACROS FOR THE REFERENCE LIST
% I.E \beginrefer \refer and \endrefer
%
\def\beginrefer{\section*{References}%
\begin{quotation}\mbox{}\par}
\def\refer#1\par{{\setlength{\parindent}{-\leftmargin}\indent#1\par}}
\def\endrefer{\end{quotation}}
%
% BEGIN THE ABSTRACT CHAPTER WITH \noindent\small, ENCLOSE IT IN A GROUP
% AND BOLDFACE THE TITLE.
%
{\noindent\small{\bf Abstract:} 
The theory of radiatively driven winds successfully explains the key points
of the stellar winds of hot massive stars.
However, there is an apparent break-down of this paradigm
at $\log L/L_{\odot} <$5.2: the stellar  wind momentum is smaller than predicted for
low luminosity early-type stars from metal poor environments,
and there are also some Galactic examples.
In this work we explore whether magnetic fields are playing a role.

}
%
% NOW COMES THE MAIN BODY OF THE ARTICLE
%
\section{Introduction}
There is an apparent breakdown of the theory of radiation driven 
winds at $\log L/L_{\odot} <$5.2,
where the wind-momentum derived from the observations
is smaller (beyond error bars) than prediced by theory.
This was first thought a metallicity effect
since it was detected in
%The theory fails to predict the wind momentum of low luminosity early-type
stars from metal-poor environments like the SMC
(Bouret et al.\ 2003, Martins et al.\ 2004).
However, there are also some  Galactic cases
(Martins et al.\ 2005, Marcolino et al.\ 2009).

We still lack an explanation but current hypothesis include a metallicity-dependent threshold luminosity
to start the wind, an early evolutionary state previous to wind onset,
decoupling of the driving ions from the plasma or magnetic fields.

We study a sample of O9-B0 young dwarfs in the Orion star forming region,
together with $\tau$~Sco, a known magnetic star.
Our first analysis of ultraviolet (UV) spectra (1150-1800\AA) with WM-basic models
revealed that all targets have smaller wind momentum than predicted
by the theory, although $\tau$~Sco's was larger and closer to
the theoretical relation. However, this analysis was not conclusive,
as the photospheric properties (\Teff~ and \logg) were adopted from
optical studies. We present here 
first results from an improved multiwavelength analysis
(from UV to the optical range) for the same sample of stars,
using CMFGEN synthetic spectra to fit the observations,
where stellar and wind parameters are determined consistently.
In this work we explore magnetic fields and the tightly 
connected wind x-ray emission as possible explanation
for the weak wind problem.

\section{Previous work: analysis with WM-basic models}

We derived the wind properties of a sample of O9-B0.5V stars from their
UV spectra. The sample includes stars in the
Orion Nebula, 10~Lac and $\tau$~Sco (Garcia \& Herrero, 2010).

The UV spectra were observed with the International Ultraviolet Explorer (IUE)
and downloaded from the INES archive
(Wamsteker et al.\ 2000).
The spectra did not display any wind features in the IUE range
with the exception of {\sc C iv}1550 (and perhaps {\sc N v}1240)
in some stars in the sample.

To analyze the UV data, 
we built for each star a grid of spherical hydrodynamical line-blanketed
non-LTE synthetic spectra
calculated with the WM-Basic code (Pauldrach et al. 2001).
WM-Basic provides a very detailed treatment of the spectral lines in the 
UV range, however, it lacks an appropriate treatment of the Stark
broadening and is not suitable for analysis of photospheric lines in the
optical range.

Consequently, the grid was run with photospheric (\Teff, \logg) parameters
derived for these stars by Sim\'on-D\'{\i}az et al. (2006)
from detailed quantitative analysis of optical spectra using 
FASTWIND (Puls et al.\ 2005).
The observed H$_{\alpha}$~ line did not exhibit a wind profile,
indicating a low mass loss rate.
The H$_{\alpha}$~ profiles of the synthetic spectra calculated
with the derived photospheric parameters were in fact insensitive
to variations of the wind parameters.
Therefore, Sim\'on-D\'{\i}az et al. could only set
upper limits for the wind-strength Q-parameter ($Q=\dot M (v_{\infty} R_{\star})^{1.5} $,
\mdot~ being the mass loss rate, \vinf~ the terminal velocity
and $\rm R_{\star}$~ the stellar radius).

Each object's customized grid has the following varying parameters:
\begin{itemize}
\item{Mass loss rate.
Starting from Sim\'on-D\'{\i}az et al. upper limit of Q as first value,
\Mdot~ was decreased until it reproduced the observed weak UV features.
}
\item{Terminal velocity.
The traditional method to derive this parameter 
could not be used,
as the UV stellar spectra did not display any saturated lines.
The values of \vinf~ considered in the grid were estimated
from the escape velocity (see Lamers et al.\ 1995), $\rm \pm$~ 500km/s.
}
\item{Shocks in the wind, 
parameterized by the X-ray luminosity released 
in the shock cooling zone. 
This parameter could not be constrained
since observations of the  {\sc O vi}1033 line in the Far-UV 
were not available for all stars.
%Chandra X-ray data exist  but igual lo que miden no es lo mismo
%que lo que hay que meter de input en WM-basic.
The adopted values for this parameter 
were: \Lx=-6.5, -7.0, -7.5, -8.0 and shocks off.}
\end{itemize}

We found that the grid models predicted strong P~Cygni profiles of
{\sc C iii}1176, {\sc C iv}1550, {\sc N v}1240 and {\sc Si iv}1398
not seen in the observations, unless mass loss rate was extremely
small. 
Moreover, the models could not reproduce
the observed strength of all these lines simultaneously.
Only upper limits for \Mdot~ could be set.

\begin{figure}[t]
\centering
\includegraphics[width=8cm]{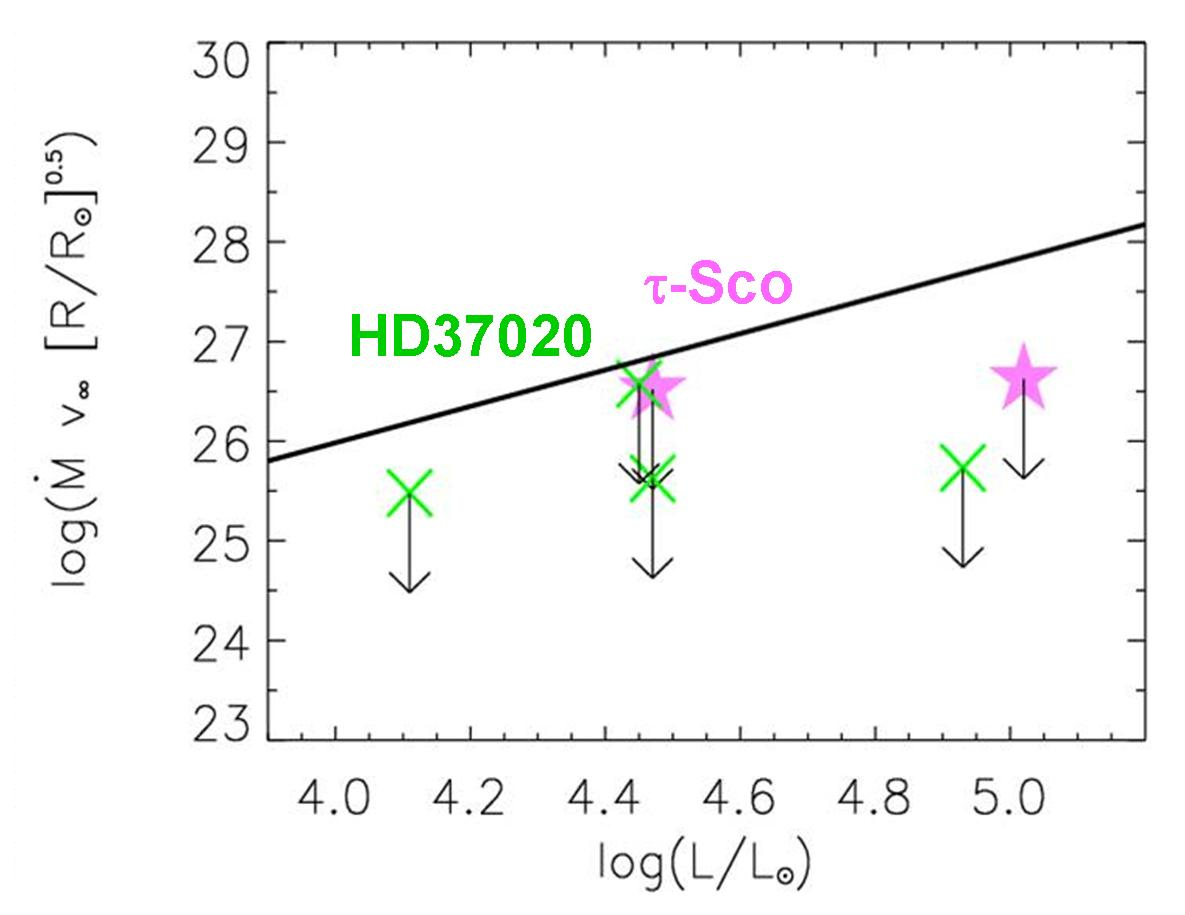}
\caption{
{\small
Wind-momentum Luminosity Relation (WLR) for the stars analyzed in
Garcia \& Herrero (2010). We include the theoretical relation of Vink et al.\ (2000)
as reference \label{fig_1}.
}
}
\end{figure}

The subsequent upper-limits for the Wind momentum-Luminosity Relation (WLR) are
systematically smaller than the theoretical prediction. Only
$\tau$~Sco and HD~37020 are close to Vink et al. (2000)'s relation (see Fig.\,\ref{fig_1}).
\textbf{$\mathbf \tau$~Sco is a known magnetic star and HD~37020 has been
recently reported a candidate based on its X-ray behaviour 
(Steltzer et al.\ 2005)
and its non-thermal radio-emission (Petr-Gotzens \& Massi\ 2007).}

\section{The interplay of magnetic fields and X-ray luminosity}

Magnetically confined wind shocks, colliding loops and reconnection events
may produce strong, hard X-ray emission in the lower wind (see for instance Puls et al.\ 2009).
This leads to increased ionization; since ions have less lines,
the radiative line-acceleration is weaker.
To reproduce the observed wind features, models with enhanced \Mdot~ are required.
This could explain why $\tau$~Sco and HD~37020 are closer to the theoretical WLR.

\subsection{A word of caution}
Schulz et al.\ (2006) found that the X-ray emission of HD~37041
may be produced as close as less than 1 stellar radius from the photosphere
(see also Cohen 2010).

X-rays in the expanding atmospheres of O stars 
alter the ionization balance of elements, shifting it towards higher values,
thus mimicking the effects of higher temperature (Garcia\ 2005).
This effect should not affect photospheric line diagnostics, nor the effective temperature
derived from photospheric lines, as shocks were thought to
be produced further out in the atmosphere.
However, if produced close enough to the photosphere, 
X-rays may impact the \Teff~ determinations
(Najarro et al. 2010 in prep.).
The derived \mdot~ would also change, as wind indicators are also sensitive to \teff.

\section{Panchromatic analysis with CMFGEN models}

The complexity of this problem requires that
all stellar (photospheric+wind) parameters
must be derived jointly and consistently,
to properly take into account the interplay of the different
contributing factors.
We have embarked on a project to re-analyze the same sample of stars,
using better observations and a more suitable model atmosphere code.

We will use the IUE UV spectra,
combined with high resolution echelle spectra
from the IACOB database 
(see S. Sim\'on-D\'{\i}az et al.'s contribution to these proceedings).
By fitting simultaneously the UV and optical lines, 
tighter constraints are set on the
temperature and velocity structure of the atmosphere,
leading to a better characterization of the 
ion-stratification and the wind.

We used the CMFGEN code (Hillier \& Miller 1998) to calculate
synthetic spectra covering from the UV to the optical range.
CMFGEN solves the spherically expanding non-LTE
atmosphere in the co-moving frame,
providing a consistent solution for the processes in the 
photosphere and the wind.
In addition, the code includes a simulation of shocks,
allowing the user to set up the
temperature of the shock region
and the wind-depth where they form.

\begin{figure}[t]
\begin{minipage}{8cm}
\centering
\includegraphics[width=8cm]{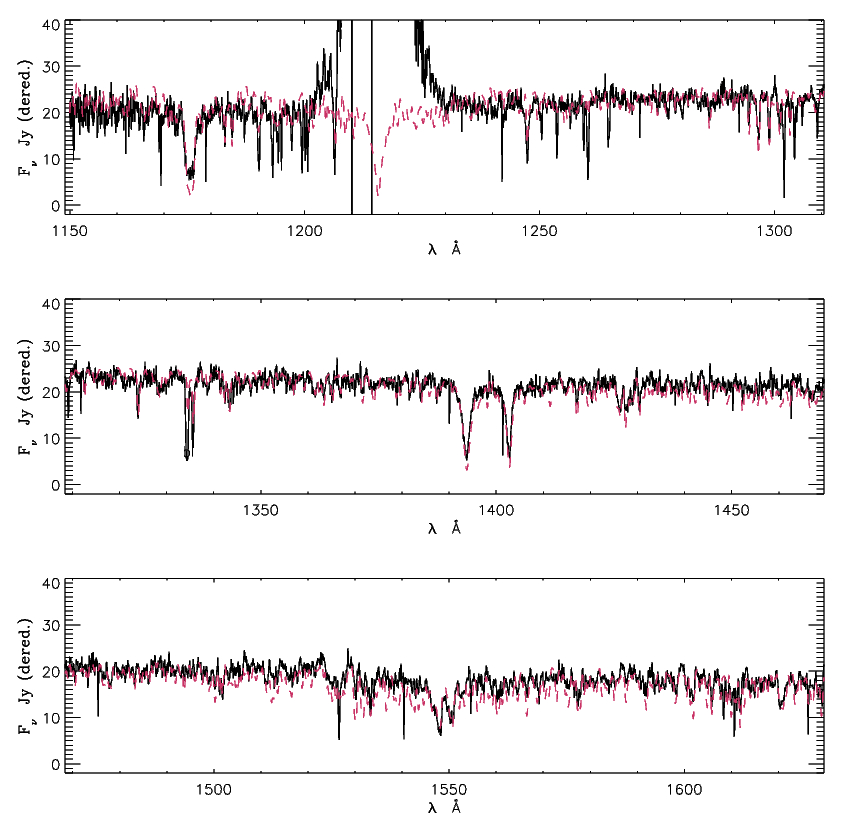}
\caption{
{\small
CMFGEN fit (red,dashed) to the IUE flux-calibrated spectrum (black, solid)
of HD~37020. The observed wind signatures are very weak.
The synthetic spectrum reproduces well the
{\sc C iii}1176 blend, {\sc Si iv}1400 and {\sc C iv}1550 lines,
and the iron-nickel continuum.
\label{fig_2}
}
}
\end{minipage}
\hfill
\begin{minipage}{8cm}
\centering
\includegraphics[width=8cm]{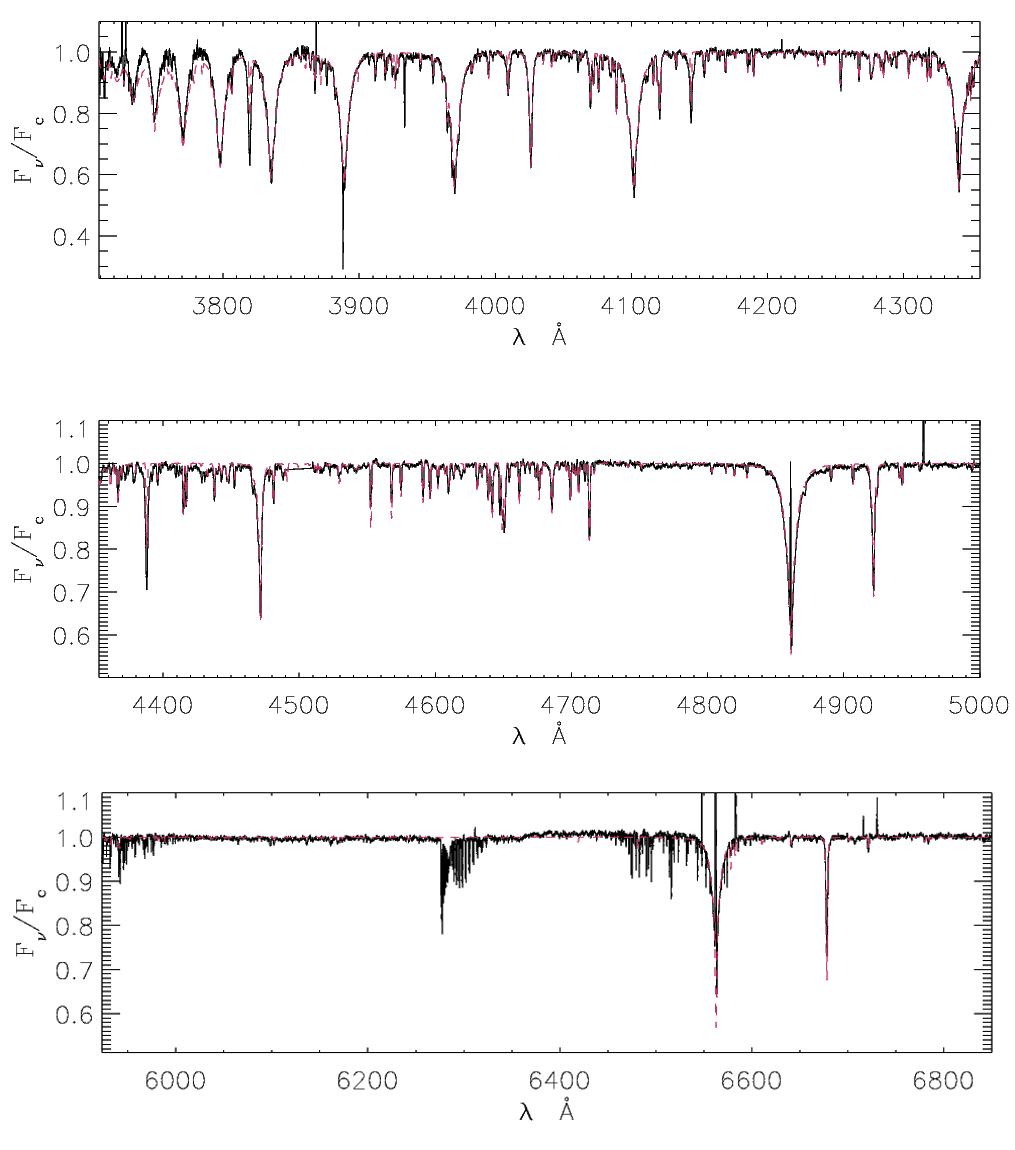}
\caption{
{\small
Same as Fig.\,\ref{fig_2}, normalized spectra in the optical range.
The CMFGEN model reproduces well the photospheric features
observed in these wavelengths.
This indicates a correct consistent determination
of photospheric and wind parameters.
H$_{\alpha}$~ (bottom panel) exhibits no clear wind feature.
\label{fig_3}
}
}
\end{minipage}
\end{figure}

We present in
Figs.\,\ref{fig_2} and \ref{fig_3} 
our preliminary fits to HD~37020.
The CMFGEN model reproduces simultaneously the photospheric and 
[weak] wind features,
and produces a good fit to the optical and UV spectra.

\section{Conclusions}

New analyses with CMFGEN models will shed new light on the problem of weak winds.
We will model the UV and optical spectra of the sample of
stars in Orion included in the high-resolution, high-S/N IACOB spectroscopic database.
New results in the X-ray domain from Chandra
(see for instance Steltzer et al.\ 2005)
will be essential.

Results on this relatively large sample, which includes young stars,
known magnetic rotators and X-ray emitters will help us
disentangle the role played by each of these factors in the weak wind problem.

%
% USE A SECTION WITHOUT NUMBER FOR THE ACKNOWLEDGEMENTS
%
\section*{Acknowledgements}
Funded by spanish MICINN under CONSOLIDER-INGENIO 2010,
program grant CSD2006-00070,
and grants AYA2007-67456-C02-01
and  AYA2008-06166-C03-01.
%
% BEGIN THE REFERENCE LIST WITH \beginrefer
% USE \refer BEFORE THE REFERENCES AND BEGIN A NEW PARAGRAPH AFTER THE 
% REFERENCE !
% DO NOT FORGET TO END THE LIST WITH \endrefer
%
\footnotesize
\beginrefer
\refer Bouret J.-C., et al., 2003, ApJ, 595, 1182

\refer Cohen D. \ 2010, in \textit{IAU Symposium 272: Active OB stars}

\refer Garcia M., 2005, \textit{PhD Thesis}

\refer Garc{\'{\i}}a, M., \& Herrero, A., 2010, in
  {\it VIII Scientific Meeting of the Spanish Astronomical Society},
  Highlights of Spanish Astrophysics V, 407

\refer Hillier, D. J., \& Miller, D. L., 1998, ApJ, 496, 407

\refer Lamers H. J. G.. L. M., et al., ApJ, 455, 269

\refer Marcolino W.L.F , et al., 2009, A\&A, 498, 837

\refer Martins F., et al., 2004, A\&A, 420, 1087

\refer Martins F., et al., 2005, A\&A, 441, 735

\refer Pauldrach, A.W.A., et al. 2001, A\&A, 375, 161

\refer Petr-Gotzens, M. G., \& Massi, M., 2007, MmSAI, 78, 362

\refer Puls J., et al., 2005, A\&A, 435, 669

\refer Puls J., et al., 2009, AIPC, 1171, 123

\refer Sim\'on-D\'{\i}az S., et al., 2006, A\&A, 448, 351

\refer Stelzer, B., et al., 2005, ApJSS, 160, 557

\refer Schulz, N. S., et al., 2006, AoJ, 656, 636

\refer Vink, J., et al., 2000, A\&A, 362, 295

\refer Wamsteker, W., et al., 2000, Ap\&SS, 273, 155

\endrefer           
\end{document}